\documentclass[preprint,superscriptaddress,prd,showpacs,aps,tightenlines]{revtex4}
\usepackage{txfonts}
\usepackage{wasysym}
\usepackage{graphicx,epsfig,amssymb}
\usepackage{multirow}
\usepackage{verbatim}

\begin{document}

\title{Probing $W^\prime_L WH$ and $W^\prime_R W H$ Interaction at LHC}
\author{Shou-Shan Bao}\email{ssbao@sdu.edu.cn}\affiliation{School of Physics, Shandong University, Jinan Shandong 250100, People's Republic of China}
\author{Hong-Lei Li}\email{lihl@mail.sdu.edu.cn}\affiliation{School of Physics, Shandong University, Jinan Shandong 250100,  People's Republic of China}
\author{Zong-Guo Si}\email{zgsi@sdu.edu.cn}\affiliation{School of Physics, Shandong University, Jinan Shandong 250100, People's Republic of China}
\affiliation{Center for High-Energy Physics, Peking University, Beijing 100871, People's Republic of China}
\author{Yu-Feng Zhou}\email{yfzhou@itp.ac.cn}\affiliation{ Kavli Institute for Theoretical Physics China (KITPC)
\\ Key Laboratory of Frontiers in Theoretical Physics, Institute of
Theoretical Physics, Chinese Academy of Science, Beijing,100190,
People's Republic of China}
\begin{abstract}
  Many new physics models predict the existence of TeV-scale charged gauge
  boson $W^\prime$ together with Higgs-boson(s). We study the $W^\prime WH$
  interaction and explore the angular distribution of charged lepton to
  distinguish $W_R^\prime WH$ from $W_L^\prime WH$ in the $pp\to HW\to b \bar b l
  \nu$ process at the LHC. It is found that a new type forward-backward asymmetry($A_{FB}$)
 relating to the angle between the direction of the charged lepton in the $W$ rest frame 
and that of the reconstructed $W^\prime$ in the laboratory frame is useful  to 
investigate the properties of $W^\prime W H$ interaction.
We analyze the standard model backgrounds  and develop a set of cuts to highlight the signal and suppress the backgrounds at LHC. We find that $A_{FB}$ can reach 0.03(-0.07) for $W_R^\prime$($W_L^\prime$) production at $\sqrt{S}=14$ TeV.
\end{abstract}
\maketitle

\newcommand{\newc}{\newcommand}
\newc{\gsim}{\lower.7ex\hbox{$\;\stackrel{\textstyle>}{\sim}\;$}}
\newc{\lsim}{\lower.7ex\hbox{$\;\stackrel{\textstyle<}{\sim}\;$}}

\def\beq{\begin{equation}}
\def\eeq{\end{equation}}
\def\beqn{\begin{eqnarray}}
\def\eeqn{\end{eqnarray}}
\def\calM{{\cal M}}
\def\calV{{\cal V}}
\def\calF{{\cal F}}
\def\half{{\textstyle{1\over 2}}}
\def\quarter{{\textstyle{1\over 4}}}
\def\ie{{\it i.e.}\/}
\def\eg{{\it e.g.}\/}
\def\etc{{\it etc}.\/}
\newcommand{\bspace}{\!\!\!}
\def\met{\mbox{$E{\bspace}/_T$}}


\def\inbar{\,\vrule height1.5ex width.4pt depth0pt}
\def\IR{\relax{\rm I\kern-.18em R}}
 \font\cmss=cmss10 \font\cmsss=cmss10 at 7pt
\def\IQ{\relax{\rm I\kern-.18em Q}}
\def\IZ{\relax\ifmmode\mathchoice
 {\hbox{\cmss Z\kern-.4em Z}}{\hbox{\cmss Z\kern-.4em Z}}
 {\lower.9pt\hbox{\cmsss Z\kern-.4em Z}}
 {\lower1.2pt\hbox{\cmsss Z\kern-.4em Z}}\else{\cmss Z\kern-.4em Z}\fi}


\section{Introduction}
Although the standard model (SM) of particles is extremely successful in
phenomenology, there are remaining problems not well understood, such as the
gauge hierarchy problem, the origins of fermion masses, mixing and $P/CP$
violation etc. The SM fails to explain the baryon-antibaryon asymmetry in the
universe and cannot provide a viable dark matter candidate.  It is commonly
believed that the SM can only be a low energy effective theory of a more
fundamental theory. There already exists various well-motivated new
physics models beyond the SM, such as the supersymmetric models
~\cite{Wess:1973kz,Haber:1984rc,Martin:1997ns}, models with extra dimensions
~\cite{Klein:1926tv,ArkaniHamed:1998rs,Randall:1999vf,ArkaniHamed:2001ca}, the
little Higgs models
~\cite{ArkaniHamed:2001nc,ArkaniHamed:2002qx,Schmaltz:2002wx,Han:2003wu} and
the left-right symmetric
models (LRSMs) ~\cite{Pati:1974yy,Mohapatra:1974hk,Mohapatra:1974gc,Senjanovic:1975rk,Mohapatra:1977mj},
etc.  Most new physics models introduce new heavy particles, such as the new
neutral ($Z^\prime$) and charged ($W^\prime$) gauge bosons, etc. The signals of
these new gauge bosons at the LHC have been extensively studied
~\cite{Olness:1984xb,Feldman:2006ce,Agashe:2007ki,Agashe:2008jb,Galloway:2009xn,Agashe:2009bb,delAguila:2010mx,Gao:2010zz}.
If the new particles beyond SM are discovered, one needs to go a step further
to know their properties such as masses and couplings to the SM particles. In a
recent analysis ~\cite{Gopalakrishna:2010xm}, it has been shown that the
chirality of the charged gauge boson to the SM fermions can be determined
by an angular distribution asymmetry of the final state leptons in the process
$pp\to W'_{L,R}\to t\bar b$ followed by $t \to b l \nu$, which is useful in
distinguishing different new physics models.

Of course, one of the primary goals of the LHC is to discover the
light Higgs-boson which is essential for testing the electroweak
symmetry breaking in the SM. In proton-proton collision, the
gluon fusion, $gg\to H$, is the dominant channel for Higgs boson production
throughout the Higgs mass range in the
  SM. Current electroweak fits, together with the LEP exclusion limit, favor a
  light Higgs around 120 GeV \cite{ALEPH:2010vi}, where $H\to b\bar{b}$ decay mode is
  dominated. However $gg\to H\to b\bar{b}$ is overwhelmed
   by the large QCD backgrounds.
   Thus the rare channel $gg\to H\to
  \gamma\gamma$ is explored to  be a golden channel for light Higgs searching at LHC
  due to the clean background. There are also other important Higgs-boson
  production processes, such as vector boson fusion and the associated production
  with $t\bar{t}$, $W^\pm$ and $Z$, etc. A detailed review can be found in
  ~\cite{Djouadi:2005gi}. Once a light Higgs boson is found, one still
needs to know if it belongs to the SM or some other new physics models, as
many new physics models contain one or more Higgs bosons which may have
different properties such as flavor changing interactions with SM fermions, or
coupling to other new particles such as the new gauge bosons $W'$ and $Z'$.

If both the new gauge bosons and the light Higgs bosons are discovered at the LHC,
investigating the possible interaction between them will shed light on the nature
of the underlying new physics. A particularly interesting interaction is the
coupling of Higgs to the new charged gauge boson $W'$ and the SM charged gauge
boson $W$.  In general, this type of coupling appears when the Higgs boson is
charged under more than one nonabelian gauge group or there exists mixing
between the Higgs bosons of different type.  The $W'WH$ coupling appears in
various models such as the extra dimension models, the little Higgs models and
the left-right symmetry models. But the nature of the $W'$ involved in the
interactions may be quite different.

The $W'WH$ coupling is of particular importance in probing the LRSM.
Unlike the extra dimension models and the little Higgs models, the
$W'$ in the LRSM couples mostly to right-handed SM fermions. The
existence of the $W'WH$ interaction  arises from the bi-doublet
nature of the Higgs boson which is essential for generating  fermion
masses in this model.

In this paper we shall focus on searching for the signal of the possible
coupling $W'WH$ at the LHC and determining the chirality of $W'$, which may
not only provide complementary information on properties of the $W'$ from
other channels such as $W'\to t b$ but also reveal the nonstandard interactions
of the light Higgs boson.
We would like to use the following process
\beqn pp \rightarrow W^\prime(W) \rightarrow H W \rightarrow b \bar b l \nu,
\eeqn  to explore the $W^\prime WH$ interaction.
 We show that the chirality of $W'$ coupling to fermions is correlated to the angular
distribution of the final charged leptons through the $W^\prime W H$ vertex. A new type of forward-backward
asymmetry determined by the angle between the direction of the charged lepton and that of the final particle
 system indicates the different properties between $W_R^\prime W H$ and $W_L^\prime W H$.

This paper is organized as follows.  In Sec. ~\ref{th-framework}, we briefly
discuss the coupling of $W^\prime W H$ from the LRSM and other models and give
the formulas for the differential cross section. The angular correlations of the final
states related to the process $q \bar q^\prime \rightarrow W^\prime(W) \rightarrow HW
\rightarrow b \bar b l \nu$ are shown as well. In Sec. ~\ref{numerical}, the numerical results of $pp\rightarrow
W(W^\prime) \rightarrow H W$ with $H\to b \bar b$, $W\to l \nu$ are presented. We finally  give a short summary  in Sec. ~\ref{summary}.
%
\section{theoretical framework}\label{th-framework}
\subsection{ $W^\prime WH$ vertex in new physic models}
The $W'WH$ vertex appears in many new physics models. As an example we first consider
the LRSM in which the $W'WH$ coupling strength is large.
In the LRSM, the gauge group is expanded to $SU(2)_L \times SU(2)_R \times
U(1)_{B-L}$, and the right-handed fermions are doublets under $SU(2)_R$.
To obtain the gauge invariant Yukawa interaction, one must introduce at least one
Higgs bi-doublet
\beqn \phi= \left( \begin{array}{cc}
    \phi_1^0 & \phi_1^+ \\ \phi_2^- & \phi_2^0 \end{array} \right),
\eeqn
which transforms as a doublet under both  $SU(2)_L$ and $SU(2)_R$. Therefore, it
couples to both the left-handed and right-handed gauge bosons $W_L$ and $W_R$.
In a version of the  minimal  LRSM~\cite{Mohapatra:1974gc,Beg:1977ti},
two higgs triplets $\Delta_{L,R}$ are introduced to break the left-right symmetry and generate the
tiny neutrino masses
\beqn \Delta_L=
\left( \begin{array}{cc}\delta_L^+/\sqrt{2}& \delta_L^{++}\\\delta_L^0&
    -\delta_L^{+}/\sqrt{2} \end{array} \right), ~~\Delta_R=
\left( \begin{array}{cc}\delta_R^+/\sqrt{2}& \delta_R^{++}\\\delta_R^0&
    -\delta_R^{+}/\sqrt{2} \end{array} \right).  \eeqn
The vacuum expectation value of the right-handed triplets $\langle
\delta_R^0 \rangle=v_R$ breaks the symmetry $SU(2)_L \times SU(2)_R \times
U(1)_{B-L}$ to $SU(2)_L\times U(1)_Y$, and the vacuum expectation value of the bidoublet
\beqn \langle \phi \rangle =
\frac{1}{\sqrt{2}} \left( \begin{array}{cc} k_1 & 0 \\ 0 &
    k_2 \end{array}\right), \eeqn
 breaks the electroweak gauge symmetry with
$k_+=\sqrt{k_1^2+k_2^2}\sim 246$ GeV. The minimal LRSM predicts that  the masses of charged gauge bosons are
\beqn
M_{1,2}^2=\frac{g^2}{4}{k_+^2+v_R^2\mp \sqrt{v_R^4+4k_1^2k_2^2}},
\eeqn
with $\tan \beta = k_1/k_2$ and mixing angle
$\tan2\zeta=-2k_1k_2/v^2_R$. Barenboim $et al.$ obtained a upper bound for the
mixing angle $|\zeta| <0.0333$ from the muon decay\cite{muondecay}. An upper
limit of $|\zeta| <0.005$ on the mixing angle is derived from semileptonic decay data by
Wolfenstein\cite{semileptonicdecay}. The limit of $\tan\beta$ can be obtained
from the following expression \beqn \tan2\zeta=-\frac{2 \tan
  \beta}{1+\tan^2\beta}\left(\frac {k_+} {v_R}\right)^2 \approx -\frac{2 \tan
  \beta}{1+\tan^2\beta} \left(\frac {m_W} {m_{W^\prime}} \right)^2.  \eeqn In the case of
$m_{W^\prime}=$1 TeV, the lower limit of $\tan\beta$ is 1.4 for $|\zeta|
<0.003$\cite{Frank:2010cj}. Thus in the LRSM the $W'$ is mostly right-handed, i.e.
$W'\approx W_R^\prime$ and $W\approx W_L$.

The couplings of $W_R^\prime$ to the quarks have the following form
\begin{equation}
\mathcal{L}=\frac{g_R}{2\sqrt{2}}V^{R}_{ij}\bar{q}_i \gamma^\mu(1+\gamma^5) q^{\prime}_j W^{\prime \mu}_{R}+ \cdots +h.c.,
\label{couplefermion}
\end{equation}
with $g_{R}$ the coupling constant and $V_{ij}^{R}$ the right-handed Cabibbo-Kobayashi-Maskawa quark-mixing matrix (CKM)
 elements. From the Higgs kinetic terms one obtains the $W^\prime W H$ coupling
\begin{equation}
g_{W_R^\prime W H} = g_L g_R k_+ \frac{\tan\beta}{(1+\tan^2\beta)}.
\end{equation}
The coupling strength for $WWH$ is $g_{WWH} = g^2 k_+ /2\simeq g m_W$.

In other models such as extra dimension models and Little Higgs models the
extra charged gauge bosons $W'$ couple to left-handed fermions. The coupling
strengths are proportional to the left-handed CKM matrix elements.
\beqn
\mathcal{L}=\frac{g_L}{2\sqrt{2}}V^{L}_{ij}\bar{q}_i \gamma^\mu(1-\gamma^5)
q^{\prime}_j W^{\prime \mu}_{L}+ \cdots +h.c..
\eeqn
We parametrized the $W_L^\prime W H$ vertex as \beqn W_{L}^{\prime \pm}
W^{\mp} H &\longrightarrow &(-i)R\,g_{W W H}\,g^{\mu \nu}, \eeqn
where $R$ is a model-dependent parameter. For simplicity, as an example, we suppose $R=\sin2\beta$,
 which provides the identical coupling $g_{W_R^\prime W H}= g_{W_L^\prime W H}$ for $W_R^\prime W H$
and $W_L^\prime W H$ vertexes. We also set $g_L=g_R=g$ and $V_{ij}^L=V_{ij}^R=V_{ij}$.

The $W^\prime$ mass is limited by both the experimental results and theoretical
analysis~\cite{Nakamura:2010zzi}. A $W^\prime$ boson with mass less than 788 GeV and 800 GeV is excluded
by CDF through the decays
$W^\prime \to l \nu$~\cite{wptolnuCDF} and $ W^\prime \to t\bar b$~\cite{wptotbCDF}. The D0 collaboration
obtains a lower bound at 1 TeV for a SM-like $W^\prime$~\cite{wptolnuD0}. A global fit result
~\cite{wpglobalfit}, considering the Fermi constant, Z-mass, etc., shows the lower $W^\prime$ mass bound about 300 GeV.
Otherwise, with reasonable fine-tuning restrictions one could obtain $M_{W^\prime}>300$ GeV~\cite{finetuning}.
As well as the low energy experiments, i.e., electron-hadron, neutrino-hadron and neutrino-electron processes
 restrict the mass of $W^\prime$ above 875 GeV~\cite{lowenergyex}. It is pointed out that the neutral current phenomena can provide limits to $W^\prime$ mass~\cite{Mohapatra:1983ae}, and it is summarized in ~\cite{Maiezza:2010ic} that the $W^\prime$ will be 2-3 TeV.  
From the $K_L-K_S$ mixing, the $W^\prime$ is limited to above 1.6 TeV~\cite{k_lk_smixing},
and is up to 2.45 TeV including CP-violation restrictions~\cite{cpvilation,cpvilation-ji}.
 The constraint from neutral K meson mass difference $\Delta m_K$ demonstrates that the $W^\prime$
mass well below 1 TeV is allowed due to a cancellation caused by a light charged Higgs boson~\cite{deltamk},
 while it is improved to 2.5 TeV and 4 TeV from $\Delta m_B$ and neutron electric dipole moment constraints
 ~\cite{deltamb,Wang:2008nk,Wang:2008zzu}.
\begin{figure}
\begin{center}
\includegraphics{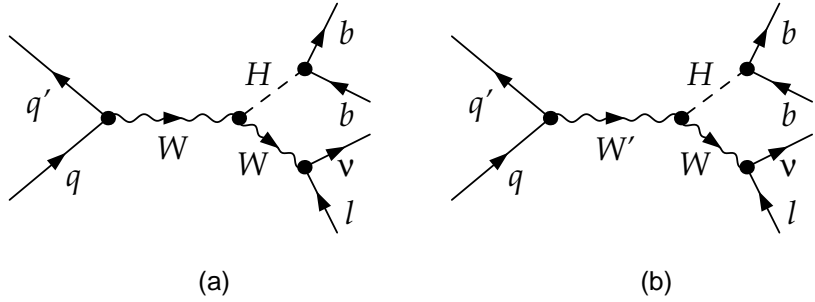}
\caption{Lowest-order Feynman diagram at the tree level for process (\ref{pro:w}) and (\ref{pro:wprime}).}
\label{fig:feyndiagram}
\end{center}
\end{figure}

\subsection{$HW$ production via $q \bar q^\prime$ annihilation}
\label{sec:WHproduction} 
\begin{figure}
\begin{center}
\resizebox{3.15in}{!}{\includegraphics*{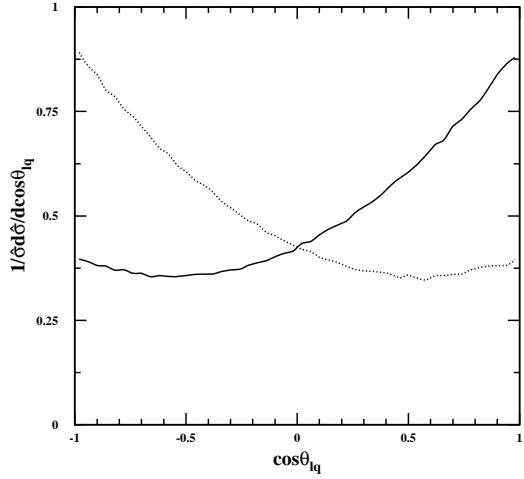}}
\caption{The angular distribution for the final charged lepton at parton level with $\sqrt{\hat{s}}=1$ TeV. The solid(dashed)
 line is the result of the left-(right-) handed $W^\prime$ with $M_{W^\prime}=1$ TeV.}
\label{fig:thetaq-nopdf}
\end{center}
\end{figure}

The search for Higgs particles is one of
the most important endeavors at LHC. Various channels can be
exploited at hadron colliders to search for a Higgs boson. In
addition to $gg\to H \to \gamma\gamma$, Higgs-boson production in
association with W or Z bosons through $q \bar q^\prime$ annihilation,
\begin{equation}
pp\to HV+X~~(H\to b \bar b, V=W ~or~ Z),
\end{equation}
is another promising discovery channel for a SM Higgs particle with mass below about 135GeV
~\cite{Nakamura:2010zzi,Butterworth:2008iy,DelFabbro:1999tf,Stange:1994bb,Kleiss:1990vca,Glashow:1978ab}.
If $W^\prime$ boson exits, it will enhance the
 cross-section around $W^\prime$ mass. In this paper we
study the properties of $W^\prime W H$ interaction via the following processes(Fig.\ref{fig:feyndiagram}),
\beqn
 q(p_q)\bar q^{\prime}(p_{\bar q})\rightarrow W^+ \rightarrow H W^+ \rightarrow b(p_b) \bar b(p_{\bar b})
   l^+(p_l)\nu(p_{\nu}),\label{pro:w}\\
 q(p_q)\bar q^{\prime}(p_{\bar q})\rightarrow W^{\prime +}\rightarrow H W^+ \rightarrow b(p_b) \bar b(p_{\bar b})
   l^+(p_l)\nu(p_{\nu}),
   \label{pro:wprime}
\eeqn
where $p_q$, $p_{\bar q}$, etc. respectively denote the 4-momentum of the corresponding particles. $H$ is a SM-like Higgs
 decaying to $b\bar b$ dominantly, thus it can be reconstructed from two b-jets at LHC. The
corresponding matrix element square averaged over the spin and color of initial partons is given by
\beqn
|\calM|^2 &=&\frac{2f_{b\bar b H}^2 |V_{p \bar q^\prime}|^2(p_b\cdot p_{\bar b})}{((s_3-m_H^2)^2+\Gamma_H^2m_H^2)}
\left\{\frac{4g^4g_{WWH}^2(p_q\cdot p_{l})
(p_{\bar q}\cdot p_{\nu})}{((s_1-m_W^2)^2+\Gamma_W^2m_W^2)((s_2-m_W^2)^2+\Gamma_W^2m_W^2)}\right.  \nonumber\\
&+& \frac {g^2g_{L(R)}^2g_{W_{L(R)}^\prime WH}^2[(1+A)^2(p_q\cdot p_{\nu})(p_{\bar q}\cdot p_{l})+
(1-A)^2(p_q\cdot p_{l})(p_{\bar q}\cdot p_{\nu})]}{((s_1-m_{W^\prime}^2)^2+
\Gamma_{W^\prime}^2m_{W^\prime}^2)((s_2-m_W^2)^2+\Gamma_W^2m_W^2)}\nonumber\\
&+&\left. \frac{2gg_Lg_{W_L^\prime WH}g_{WWH}(1-A)^2(p_q\cdot p_{l})(p_{\bar q}\cdot p_{\nu})\left[(s_1-m_{W^\prime}^2)
(s_1-m_{W}^2)+\Gamma_{W^\prime}m_{W^\prime}\Gamma_W m_W \right]}{((s_1-m_{W^\prime}^2)^2+
\Gamma_{W^\prime}^2m_{W^\prime}^2)((s_1-m_{W}^2)+\Gamma_W^2 m_W^2)((s_2-m_{W}^2)+\Gamma_W^2 m_W^2)} \right\}, \nonumber\\
\label{eq:msquare2}
\eeqn
where $s_1=\hat{s}=2p_q\cdot p_{\bar{q}}$, $s_2=2p_l\cdot p_\nu$, $s_3=2p_b\cdot p_{\bar{b}}$, $f_{b\bar b H}$ is the
Yukawa coupling of $b \bar b H$ interaction, and $V_{p \bar q^\prime}$ is the CKM matrix element. $\Gamma_H$, $\Gamma_W$
 and $\Gamma_{W^\prime}$denote the Higgs, $W$ and $W^\prime$ width, respectively, and the$W^\prime$ width is listed in the
 Appendix. The first two terms in Eq.(\ref{eq:msquare2}) stand for the matrix element square for process (\ref{pro:w}) and
  (\ref{pro:wprime}) respectively, and the third term is their interference term. $A=$1(-1) stands for right-(left-)handed
  $W^\prime$. Obviously the interference term disappears for the case of right-handed $W^\prime$ production. The
 cross section at parton level can be written as
\beqn \hat
\sigma (\hat s)=\int\frac {|\calM|^2}{2\hat
s}(2\pi)^4\delta^{(4)}\left(\sum_f p_f-p_q-p_{q^\prime}\right)
\prod_f \frac{d^3 p_f}{(2\pi)^3 2E_f},
\eeqn
where f $=$ b, $\bar b$, $l^+$, $\nu$. From the second term in Eq.(\ref{eq:msquare2}), one can notice that the angular
distribution of the charged lepton is different for the left- and right-handed $W^\prime$ bosons. In order to show this
 point clearly, we define the angle between the 3-momentum ${\bf p}_l^*$ of the charged lepton in the $W$ rest frame, and
 that(${\bf p}_q$) of the initial quark in the $ q \bar q^\prime $ center of mass system as
\beqn
cos\theta_{lq} =\frac {{\bf p}^*_l\cdot{\bf p}_q}{|{\bf p}^*_l|\cdot|{\bf p}_q|}.
\eeqn
 The differential distribution $1/\hat{\sigma}d\hat{\sigma}/d\cos\theta_{lq}$ for the partonic process
  $q \bar q^\prime \rightarrow W^\prime \rightarrow HW \rightarrow b\bar b l^+\nu $ at $\sqrt{\hat s}=$1 TeV is
  displayed in Fig.\ref{fig:thetaq-nopdf}.
Obviously, the charged leptons, produced through $W_L^\prime$($W_R^\prime$), tend to move along the direction
of the initial antiquark(quark), i.e., the $W_R^\prime W H$ and $W_L^\prime W H$ interaction may be
distinguished from this kind of angular distribution.

%

\section{numerical results}\label{numerical}
For the processes
\beqn
pp\rightarrow W^{\prime +}(W^+)\rightarrow H W^+ \rightarrow b \bar b l^+ \nu,
\label{pro:hadr}
\eeqn
the total cross section can be expressed as
 \begin{equation}
\sigma=\int dx \int dy q_i(x) \bar{q}_j(y) \hat{\sigma}(\hat{s})
\end{equation}
where $q(x)$($\bar{q}(y)$) is the parton distribution function of quark(antiquark).
CTEQ6l1 ~\cite{cteq} is used in this work. To obtain the numerical results we adopt the parameters limited in the LRSM
framework related to the $W_R^\prime$ production. For simplicity we use the same values for $W_L^\prime$ production.
\begin{figure}
\begin{center}
\begin{tabular}{cc}
\resizebox{3.15in}{!}{\includegraphics*{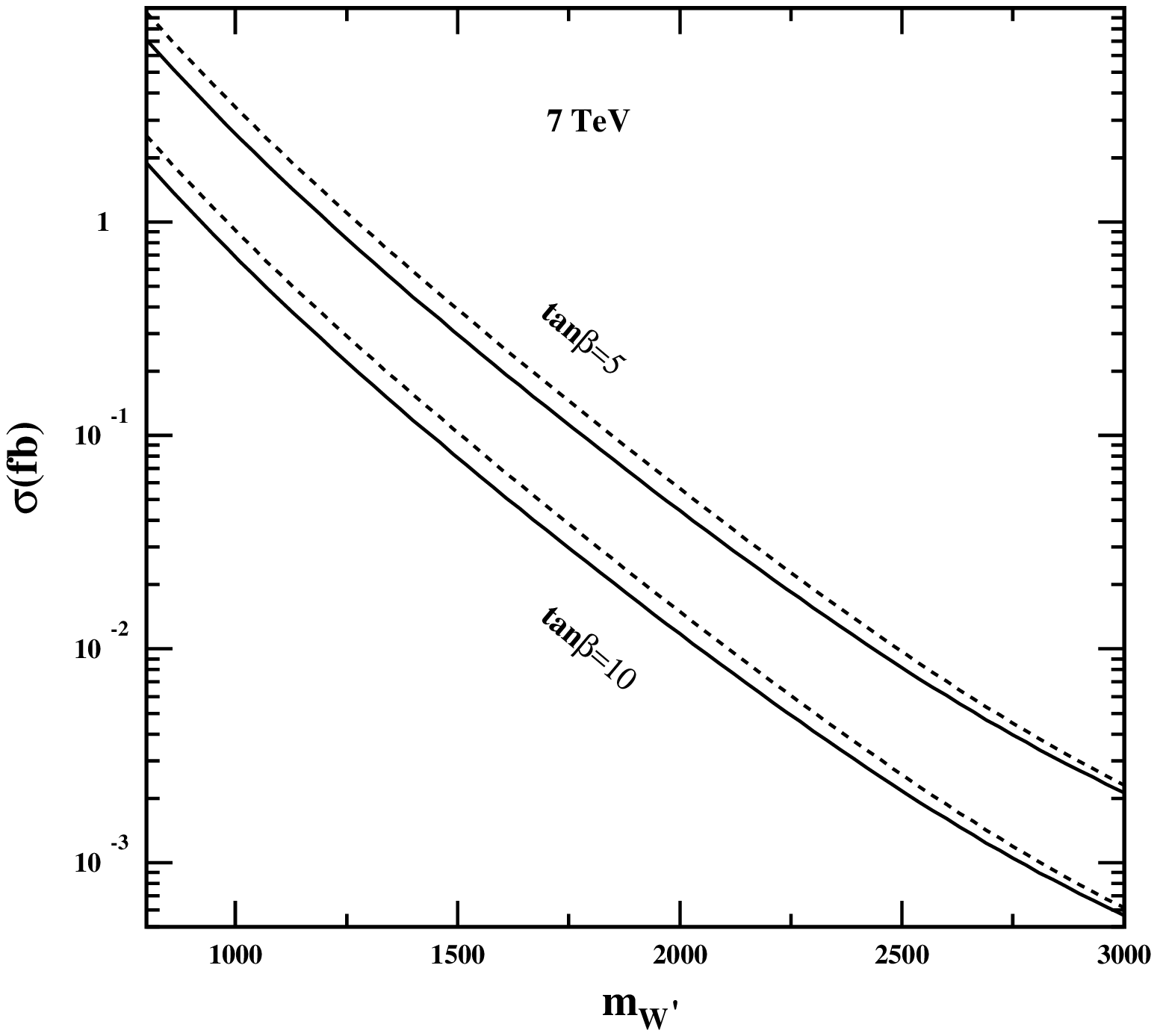}}&
\resizebox{3.15in}{!}{\includegraphics*{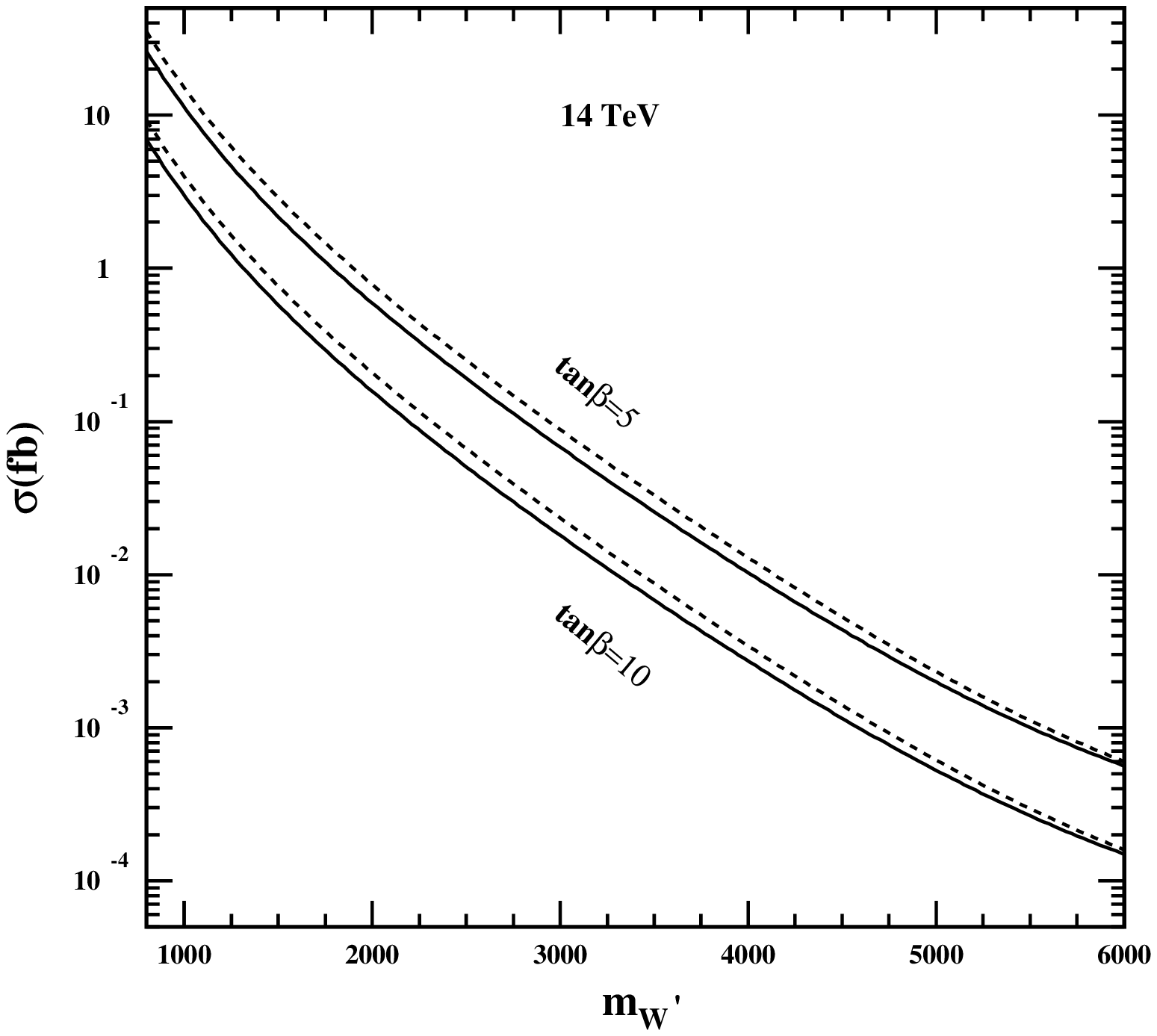}}\\
(a)&(b)\\
\end{tabular}
\caption{The total cross section distribution for process $pp\rightarrow W^{\prime +}\rightarrow H W^+ \rightarrow b\bar b l^+ \nu$ with $m_W^\prime$ at LHC for (a) $\sqrt{s}=7$ TeV and (b) $\sqrt{s}=14$ TeV. The solid (dashed)lines stand for $W_L^\prime$($W_R^\prime$).}
\label{fig:sigma-m}
\end{center}
\end{figure}
\begin{figure}
\begin{center}
\resizebox{3.15in}{!}{\includegraphics*{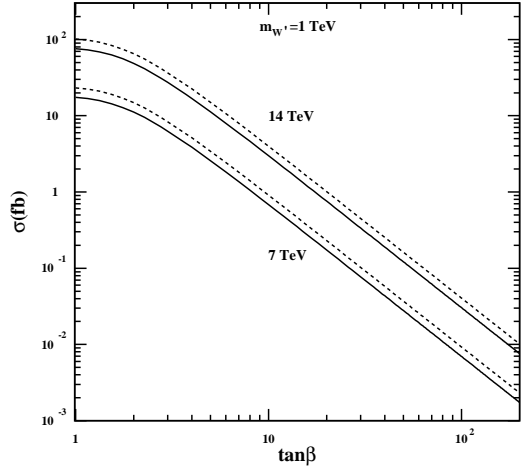}}
\caption{The total cross section distribution for process $pp\rightarrow W^{\prime +}\rightarrow H W^+\rightarrow b\bar b l^+ \nu$ with $\tan\beta$ at LHC. The solid (dashed)lines stand for $W_L^\prime$($W_R^\prime$).}
\label{fig:sigma-tanb}
\end{center}
\end{figure}
\begin{figure}
\begin{center}
\begin{tabular}{cc}
\resizebox{3.15in}{!}{\includegraphics*{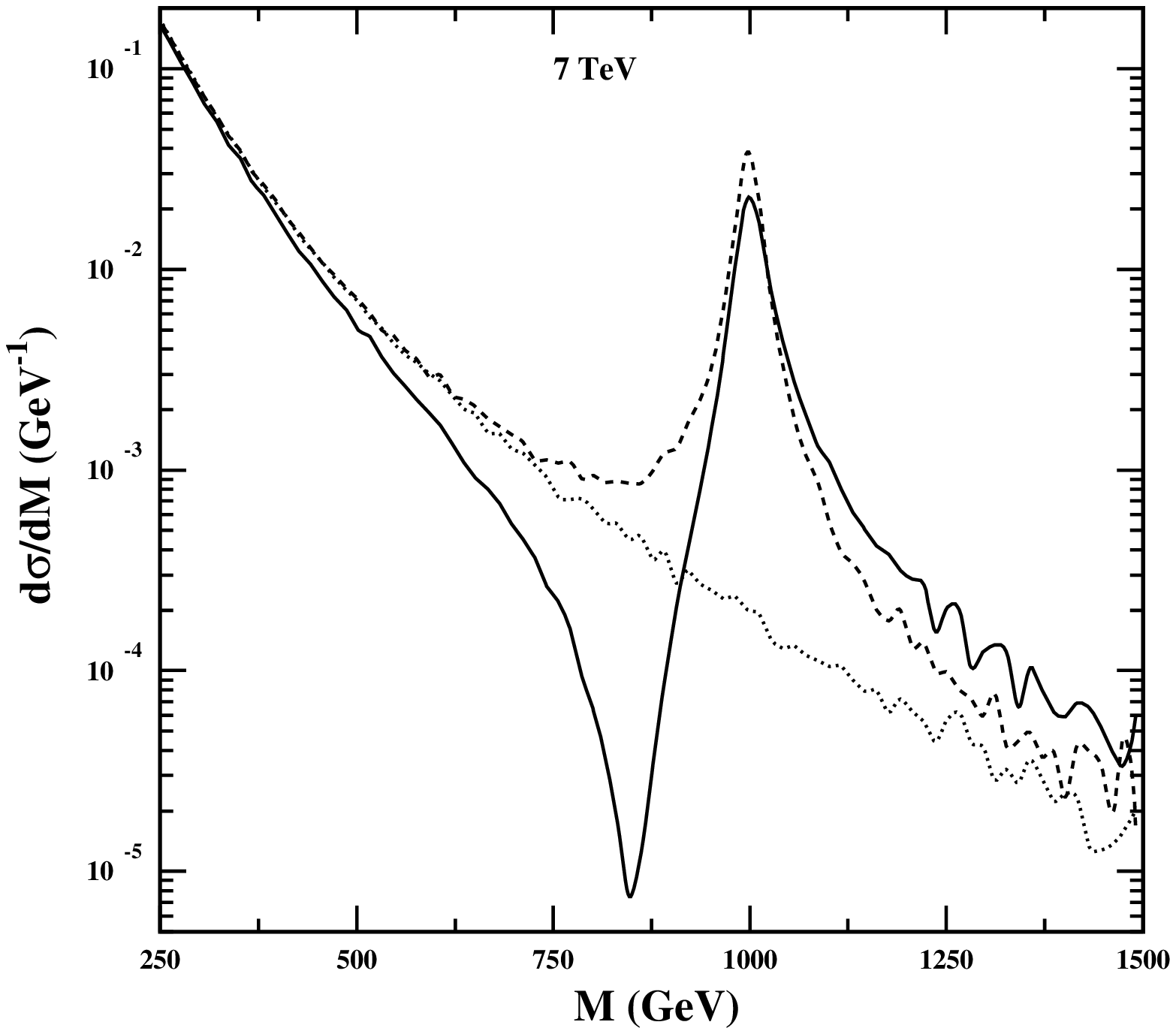}}&
\resizebox{3.15in}{!}{\includegraphics*{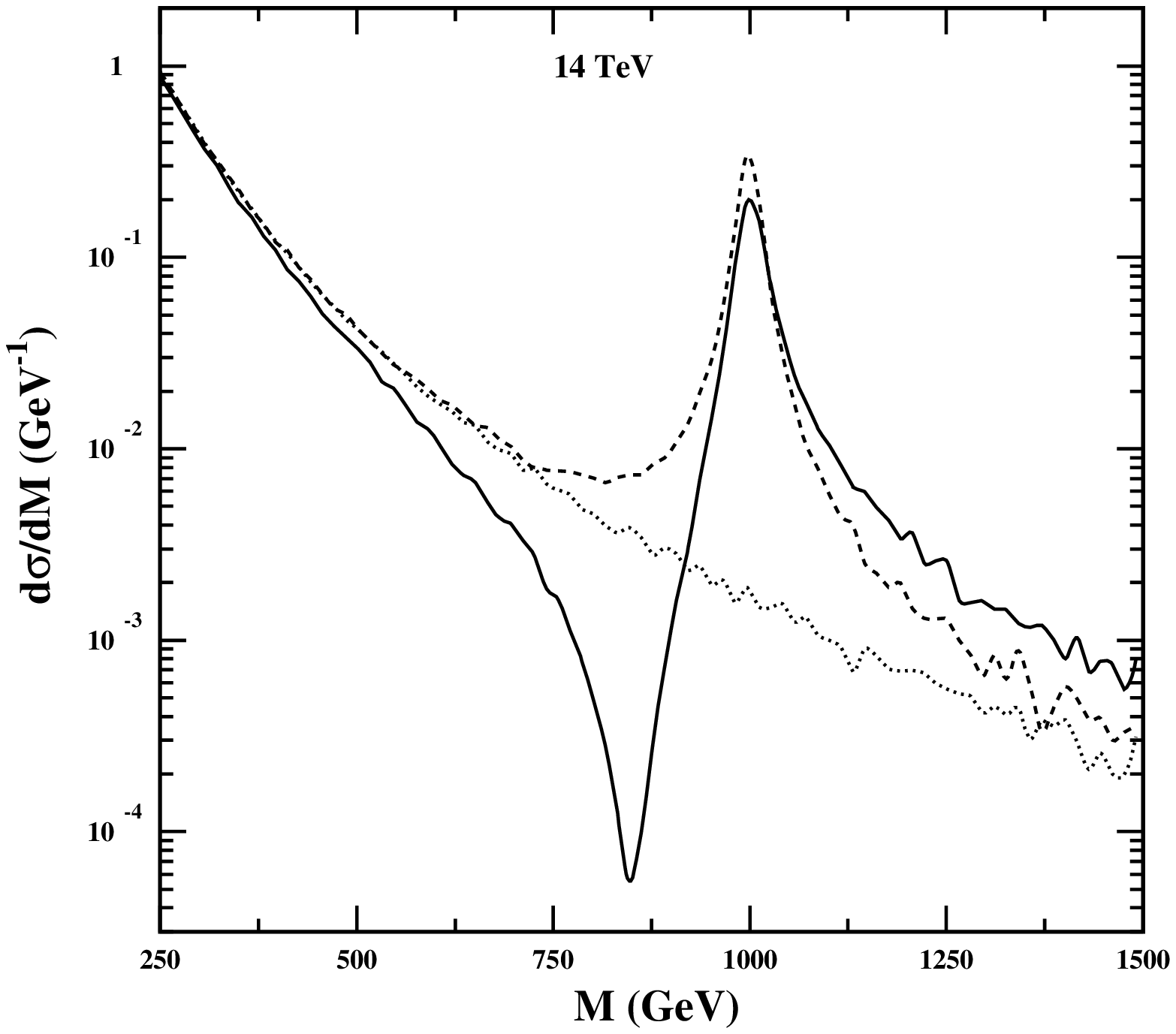}}\\
(a)&(b)\\
\end{tabular}
\caption{The differential distribution with $M$ for process $pp\rightarrow H W^+ \rightarrow b \bar b l^+\nu$. The solid (dashed) lines are contributed by $W_L^\prime+ SM$ ($W_R^\prime+ SM$), and  the dotted lines are only the results of SM.}
\label{fig:sigma-shat}
\end{center}
\end{figure}

 The total cross section for
\beqn
pp\rightarrow W^{\prime+} \rightarrow H W^+ \rightarrow b\bar b l^+ \nu
\label{pro:ppwprime}
\eeqn
at LHC versus $m_{W^\prime}$ is shown in Fig.\ref{fig:sigma-m}. With a luminosity of 100 fb$^{-1}$ at LHC, A $W^\prime$
boson production could be detected with mass up to 2(3) TeV if $\tan\beta=$10, and up to 2.5(4) TeV if $\tan\beta=$5. The
discrepancy between $W_R^\prime$ and $W_L^\prime$ is due to the different total decay widths. The cross section
 related to $\tan\beta$ for $m_{W^\prime}=$1 TeV is displayed in Fig.\ref{fig:sigma-tanb}. It is found that up
 to $\tan\beta=$70, the process (\ref{pro:ppwprime}) might be observed with luminosity of 100 fb$^{-1}$ at LHC. In
  our following numerical studies, we set $\tan\beta=$5 and $m_{W^\prime}=$1 TeV.

 Figure\ref{fig:sigma-shat} shows the differential distribution $d\sigma/dM$ of process (\ref{pro:hadr}), where
\beqn
M = \sqrt{(p_b+p_{\bar b}+ p_{l^+} +p_\nu)^2}.
\label{wprimemass}
\eeqn
 The $W^\prime$ production induces a resonance peak around the $W^\prime$ mass threshold.
For the $W_R^\prime$ production, the interference between the $W^\prime$ and $W$ bosons is zero, while for the $W_L^\prime$
the interference term (Eq.(\ref{eq:msquare2})) is negative in the region of $m_W<M<M_W^\prime$ which causes a dip in
the curve and inversely a positive   enhancement to the cross section for the case of $M_R>M_{W^\prime}$.
 This discrepancy can provide some useful information to distinguish the $W_L^\prime WH$ from  $W_R^\prime W H$.

\begin{figure}
\begin{center}
\begin{tabular}{cc}
\resizebox{3.15in}{!}{\includegraphics*{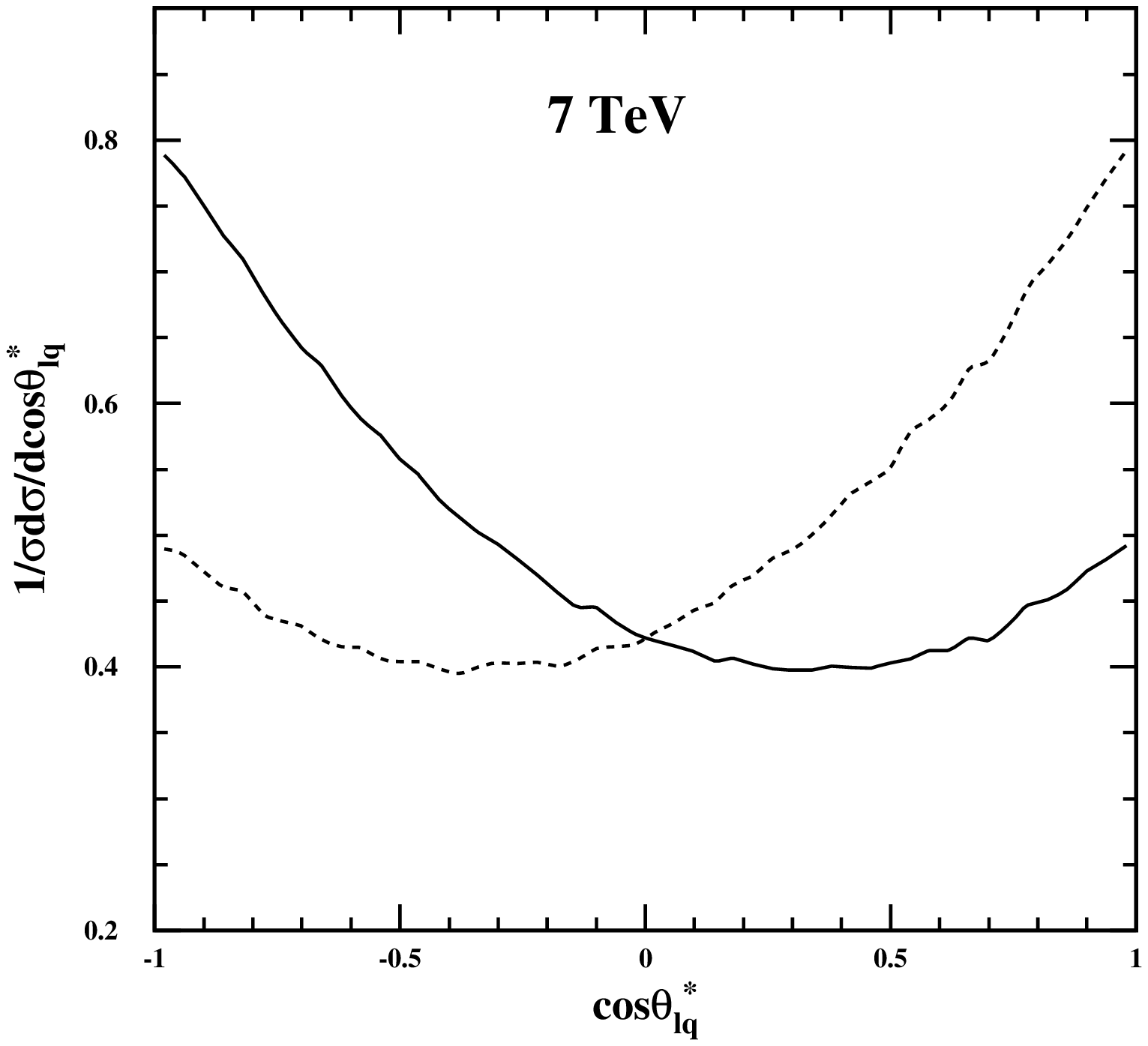}}&
\resizebox{3.15in}{!}{\includegraphics*{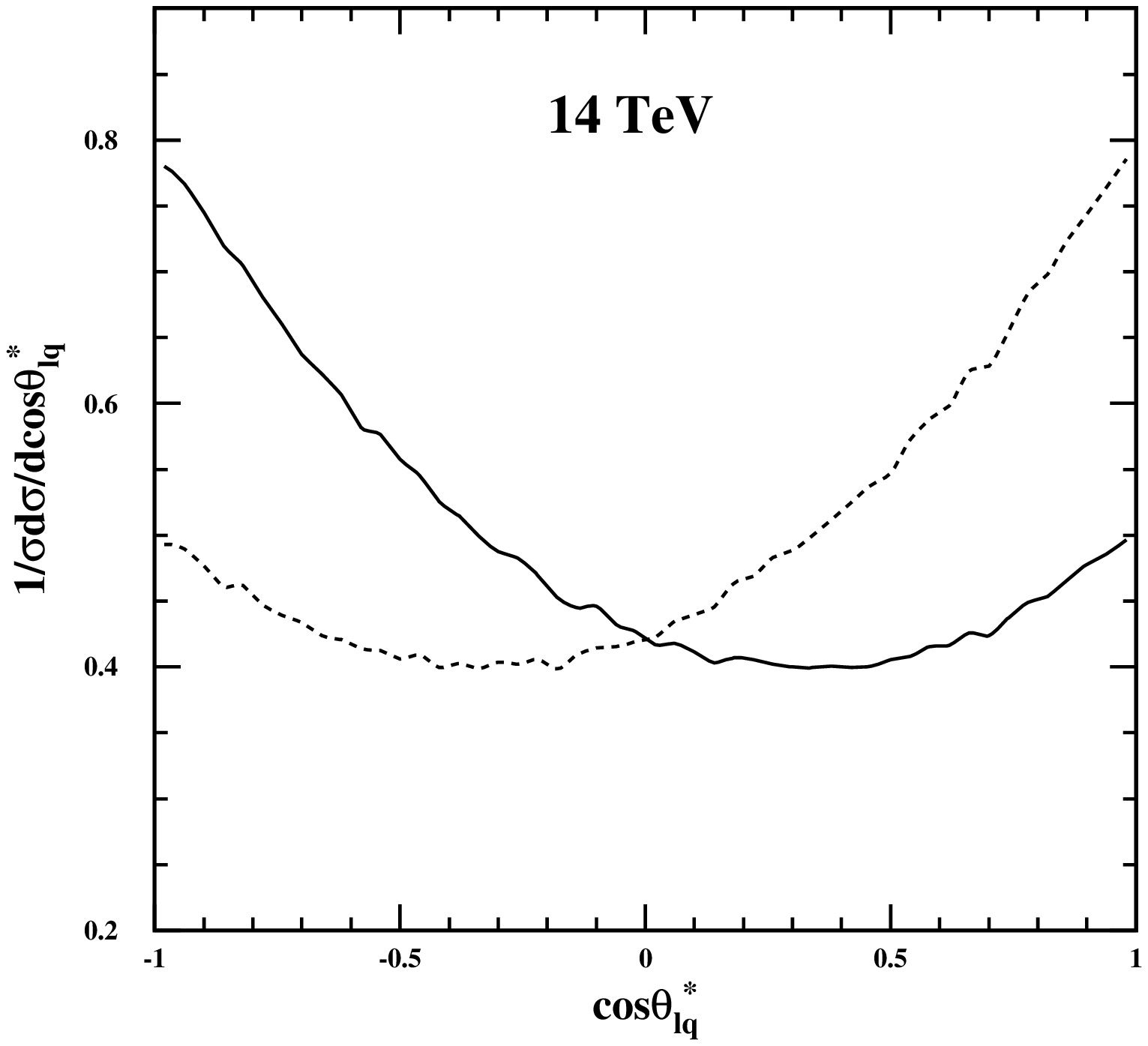}}\\
(a) &(b)\\
\end{tabular}
\caption{
The angular distribution of the charged lepton for process (\ref{pro:ppwprime}) with $\sqrt{s}=7$ TeV and 14 TeV. The solid (dashed) lines are the results for $W_L^\prime$($W_R^\prime$).}
\label{fig:theta-scut}
\end{center}
\end{figure}

Following the analysis in Sec.~\ref{sec:WHproduction}, we begin to investigate the angular distribution of charged
leptons at hadronic level. Since the LHC is a proton-proton collider, the quark can identically come from either proton, and the
charged lepton angular distribution will be symmetrized, unless we distinguish the direction of the quark from that of
 the antiquark. It can be achieved approximately based on the argument that an initial quark takes a larger momentum
  fraction than an initial antiquark on average, since the former is a valence quark in the proton and the latter a
  sea quark. Hence the final particle system ($b \bar b l^+ \nu$) will move along with the initial quark with a
  large probability. This means one can define the  total momentum of final particle system
   ${\bf p}={\bf p_b}+{\bf p_{\bar b}}+{\bf p_{l^+}}+{\bf p_{\nu}}$ instead of the quark's to redefining the charged
   lepton angular distribution,
\begin{equation}
\cos\theta^* =\frac {{\bf p}^*_l\cdot{\bf p}}{|{\bf p}^*_l|\cdot|{\bf p}|}.
\end{equation}
The differential distribution $1/\sigma d\sigma /d\cos\theta^*$ for the process (\ref{pro:ppwprime}) with $|{\bf p}|\neq 0$ is displayed in Fig.\ref{fig:theta-scut}. It is found 
that the distributions corresponding to $W_R^{\prime}$ and $W^{\prime}_L$ production
have different behavior which may be used to discriminate the 
$W_R^{\prime}WH$ and $W^{\prime}_LWH$ interaction.

\begin{figure}
\begin{center}
\begin{tabular}{cc}
\resizebox{3.15in}{!}{\includegraphics*{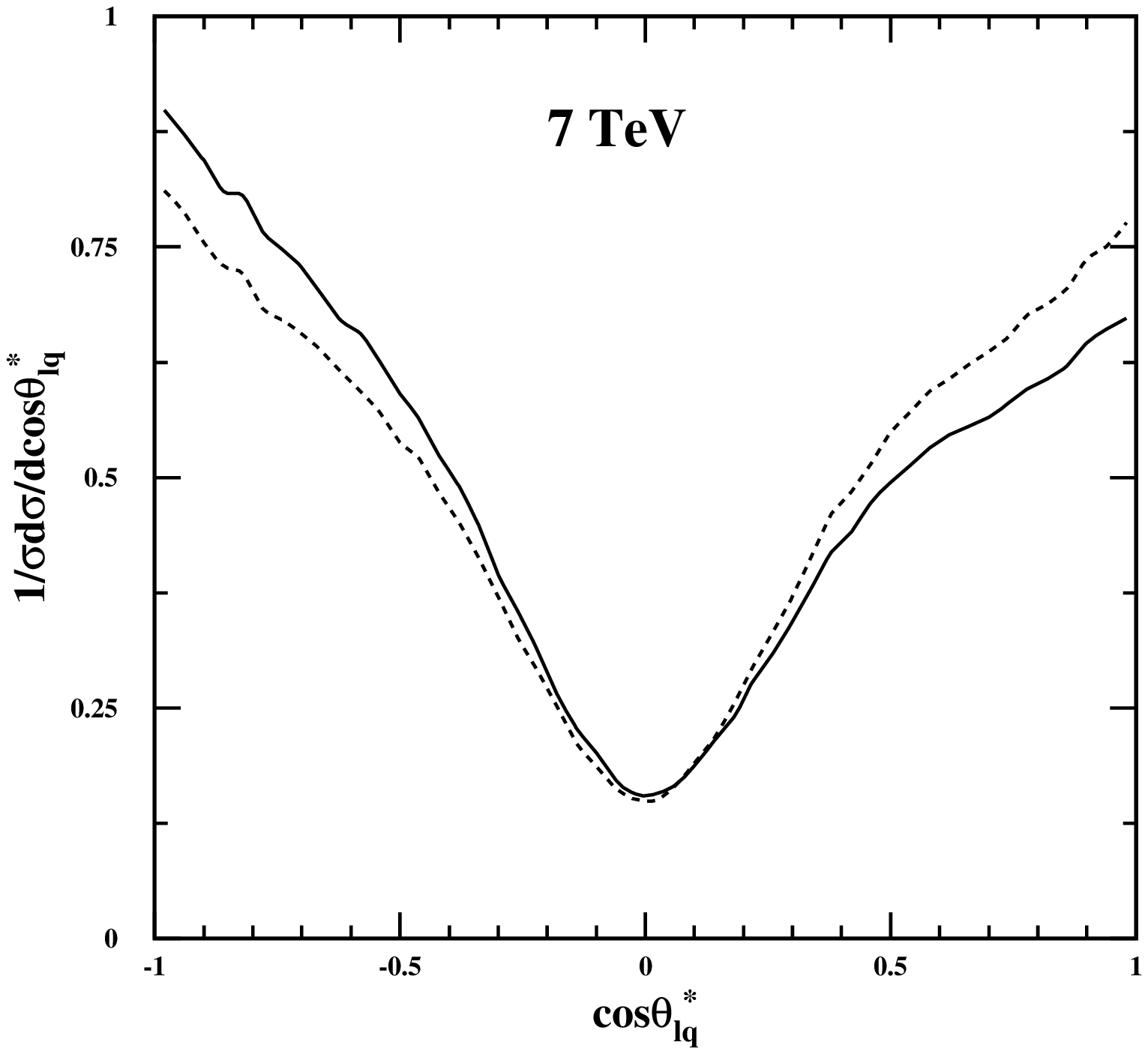}}&
\resizebox{3.15in}{!}{\includegraphics*{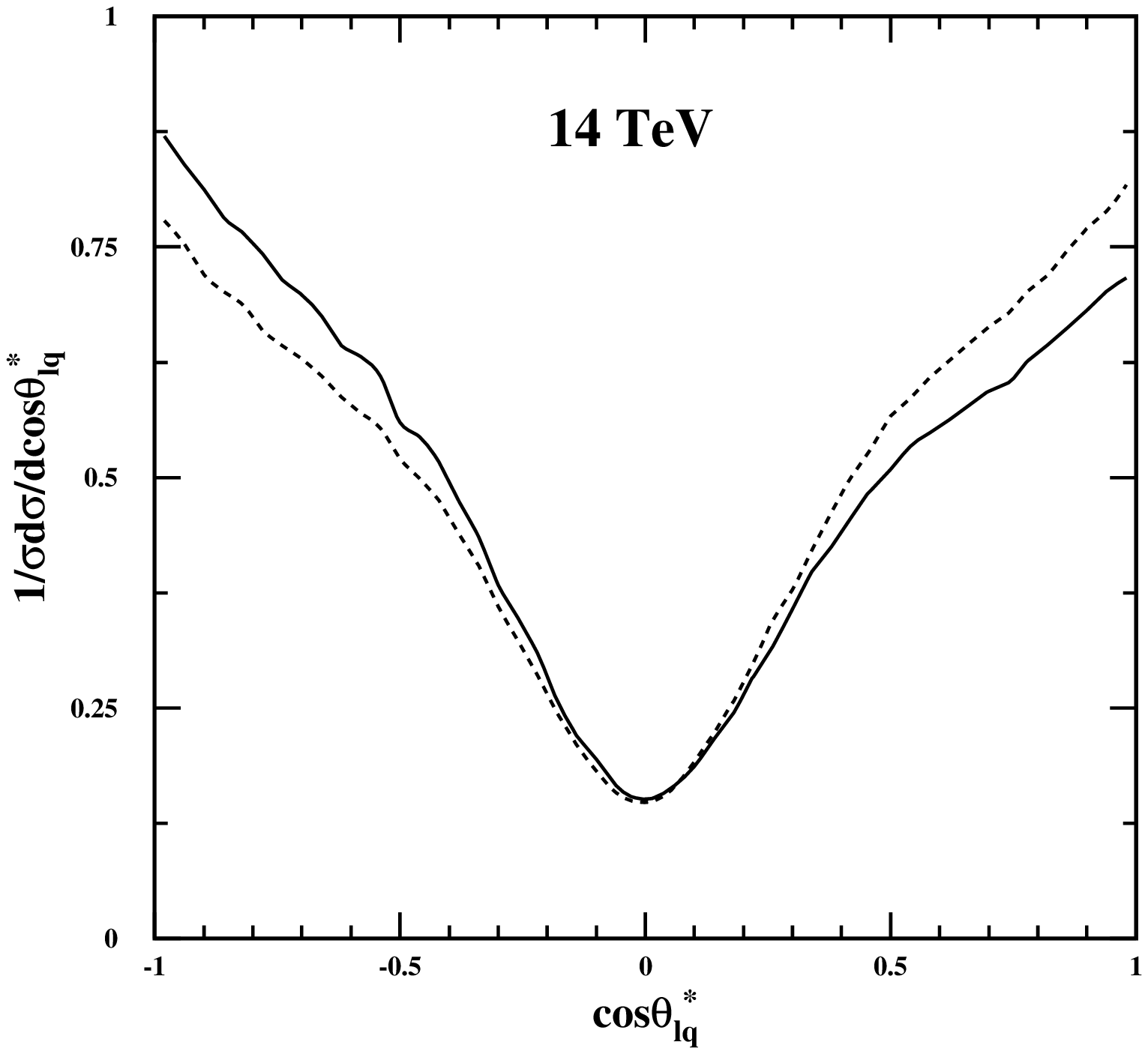}}\\
(a)  &(b)
\end{tabular}
\caption{The angular distribution of charged lepton for process (\ref{pro:hadr}) after all cuts with $\sqrt{s}=7$ TeV and 14 TeV. The solid(dashed) lines are the results for $W_L^\prime$($W_R^\prime$).}
\label{fig:etacut}
\end{center}
\end{figure}

To determine the $W^\prime$ chiral coupling from the angular distribution, one must consider
the momentum of the final states ($b\bar{b}l\nu$). To be as realistic as possible, we simulate the detector performance by smearing the lepton and b($\bar b$) quark energies according to the assumed Gaussian resolution parametrization
\begin{equation}
\frac {\sigma(E)}{E} = \frac{a} {\sqrt E} \oplus b,
\end{equation}
where $\sigma(E)/E$ is the energy resolution, $a$ is a sampling term, $b$ is a constant term, and $\oplus$ denotes a sum in quadrature. We take $a=5\% $, $b=0.55\%$ for leptons and $a=100\% $, $b=5\%$ for jets respectively\cite{Aad:2009wy}.
 Since the neutrino is an unobservable particle, one has to utilize kinematical constraints to reconstruct its 4-momentum. Its transverse momentum can be obtained by  momentum conservation from the observed particles
\begin{equation}
{\bf p}_{\nu T}=-({\bf p}_{lT}+{\bf p}_{b T} +{\bf p}_{\bar b T}),
\label{transverse}
 \end{equation}
but the longitudinal momentum can not be determined in this way due to the unknown boost of the partonic c.m. system. Alternatively, it can be solved with two-fold ambiguity through the on shell condition for the W-boson
  \begin{equation}
m_W^2=(p_{\nu}+p_l)^2.
 \end{equation}
 Furthermore one can impose the on-shell condition for the $W^\prime$-boson to remove the ambiguity.  For each possibility we evaluate the total invariant mass $M$ as defined in
Eq.(\ref{wprimemass}) and pick up the solution which is closest to the $W^\prime$ mass. With such a solution, one can reconstruct the 4-momentum of the neutrino.

In our following numerical calculations, we apply the basic acceptance 
cuts(referred as cut I)
\beqn
&p_T(l) > 50~ GeV, &~~|\eta (l)| < 2.5,  ~~p_T(j) > 50 ~GeV, ~~|\eta (j)| < 3.0, ~~\met > 50 ~GeV,\nonumber \\
& |y_c|>0.1,  & 
\eeqn
where $y_c$ is the rapidity of the reconstructed $W^\prime$ in the laboratory frame.

To purify the signal, we adopt 
$|M-M_{W^\prime}| < 100$~ GeV and $|M_{b \bar b}-M_{H}|<10$~ GeV as further cuts (referred to as cut II), where $M_{b \bar b}$ is  the invariant mass of $b \bar b$.

In Fig.\ref{fig:etacut}, we display the normalized differential distribution 
$1/\sigma d\sigma/d\cos \theta^*$ with all above cuts. Though the neutrino reconstruction
may reduce the difference of the angular distribution 
between the $W_L^\prime$ and $W_R^\prime$ production processes,
the discrepancy still exists.
In order to explore this kind of discrepancy to discriminate the $W_R^\prime W H$ 
and $W_L^\prime W H$ interaction, we define a forward-backward asymmetry as follows
\beqn
A_{FB}=\frac {\sigma(cos\theta^* \geq 0)-\sigma(cos\theta^*<0)}{\sigma(cos\theta^* \geq 0)+\sigma(cos\theta^*<0)}.
\eeqn
The total cross section together with $A_{FB}$ for process (\ref{pro:hadr}) are listed in Table.\ref{Tab:signal} at $\sqrt{S}=7(14)$ TeV.
It is found that it is possible
to distinguish $W_R^\prime$ from $W_L^\prime$ with cuts. If the 
luminosity can be accumulated to 300 $fb^{-1}$ at $\sqrt{S}=14$ TeV, about 1500 events may be found, and $A_{FB}$ can reach 0.03(-0.07)
 for $W_R^\prime$($W_L^\prime$).

\begin{table}
\begin{tabular}{|c|c|c|c|c|c|c|c|c|} \hline
&\multicolumn{4}{c|} {7 TeV}& \multicolumn{4}{c|} {14 TeV} \\
\cline{2-5} \cline{6-9}
&\multicolumn{2}{c|} {no cut}& \multicolumn{2}{c|} {cut I+II}&\multicolumn{2}{c|} {no cut}
&\multicolumn{2}{c|} {cut I+II} \\
\cline{2-3} \cline{4-5} \cline{6-7} \cline{8-9} &$W_L^\prime$&$W_R^\prime$&$W_L^\prime$&$W_R^\prime$&
$W_L^\prime$&$W_R^\prime$&$W_L^\prime$&$W_R^\prime$\\ \hline
$\sigma(fb)$ & 45.9&50.3 &0.97 &1.38 &129  &  143&4.53 &6.30  \\ \hline
$A_{FB}$     &-0.38&-0.35&-0.10&-0.01&-0.35&-0.31&-0.07&0.03  \\ \hline
\end{tabular}
\caption{The total cross-section and the forward-backward asymmetry before and after cuts with $\sqrt{S}=7$TeV and 14TeV at the LHC.}
\label{Tab:signal}
\end{table}

Finally, we consider the dominant backgrounds for our signal, i.e., 
$Wbb$, $WZ$ and $t\bar b$ \footnote{Here we do not consider the 
$pp\rightarrow W^{\prime +}\rightarrow t \bar b$ process which is 
investigated in detail in Ref.\cite{Gopalakrishna:2010xm}.
Especially, this process can be largely vetoed by reconstructing the top quark.}.
The Madgraph \cite{Alwall:2007st} software package is used 
in our simulation. The cross sections after each cut are listed 
in table.\ref{Tab:background}. Obviously, after all cuts, 
the total cross section of the dominant backgrounds is much lower than that of the signal.
 
\begin{table}
\begin{tabular}{|c|c|c|c|c|} \hline
&\multicolumn{2}{c|} {7 TeV}& \multicolumn{2}{c|} {14 TeV} \\
\cline{2-3} \cline{4-5}
&cut I&cut I+II&cut I&cut I+II\\ \hline
t$\bar b$   &0.61&0.005&11.2& 0.01\\ \hline
$W b\bar b$ &23.9&0.04 &63.8&0.08\\ \hline
WZ          &6.69&0.01 &16.6& 0.02\\ \hline
\end{tabular}
\caption{The total background cross-section after cuts with $\sqrt{S}=7$TeV and 14TeV at the LHC(unit of $fb$).}
\label{Tab:background}
\end{table}

%
\section{Summary}\label{summary}
Many theories beyond SM predict the existence of new heavy charged gauge boson $W^\prime$ and searching for Higgs boson
 at LHC motivates us to investigate the $W^\prime WH$ interaction.
In order to understand its properties, we study the process of
$pp\rightarrow W^{\prime +}(W^+)\rightarrow HW^+ \rightarrow b \bar b l^+ \nu$ in this work.
because of the resonance effect of the intermediate $W^\prime$, there appears a peak in the invariant mass spectrum of
 the final states, and for the $W_L^\prime$ production, a dip appears in the region of $m_W<M<M_W^\prime$ induced
  by the interference term. Our numerical results reveal that the angular distribution $d\sigma/d\cos\theta^*$ and
  the forward-backward asymmetry $A_{FB}$ can provide helpful information for the $W^\prime W H$ interaction.
  It is found that for $m_{W^\prime}=$1 TeV, $A_{FB}$ can reach about 0.03(-0.07) for $W_R^\prime W H$($W_L^\prime W H$)
  at LHC($\sqrt{S}=14$TeV). The backgrounds are estimated and largely suppressed by the kinematical constraints.
 Once the $W^\prime\rightarrow HW$ process is observed and enough number of events are accumulated, our
  method can be used to study the $W^\prime W H$ interaction and discriminate $W_R^\prime W H$ from $W_L^\prime W H$
   so that it is possible to distinguish different new physics models including $W^\prime W H$ interaction.
\begin{acknowledgments}
This work is supported in part by the NSFC and the Natural Science Foundation of Shandong Province (JQ200902).
\end{acknowledgments}

\appendix
\section{$W^\prime$ decay width}
For estimating the cross section of $q^\prime \bar q\rightarrow W_{L(R)}^\prime \rightarrow H W$,
a narrow width approximation is used.
The decay width of $W^\prime$ is given in the following parts. In this LRSM we have forbidden $W^\prime$
from decaying into heavy right-handed neutrinos.
The width for $W^\prime$ decaying to a pair of quarks is
\beqn
\Gamma(W_{L(R)}^\prime\rightarrow q\bar q^\prime) &=& \frac {m_{W^\prime}}{16\pi}|V_{q\bar q^\prime}|^2 g^2 g_{L(R)}^2 ,\\
\Gamma(W_{L(R)}^\prime \rightarrow t\bar b) &=& \frac {m_{W^\prime}}{16\pi}|V_{q\bar q^\prime}|^2g^2g_{L(R)}^2
(1-\frac {m_t^2}{m_{W^\prime}^2})(1-\frac {m_t^2}{2m_{W^\prime}^2}-\frac {m_t^4}{2m_{W^\prime}^4}).
\eeqn
The width for $W^\prime$ decaying to W-boson and Higgs is
\beqn
\Gamma(W_R^\prime\rightarrow  H W)=\frac {g_{W^\prime W \phi}^2}{24\pi m_{W^\prime}^2}p_f
[-6+\frac {(m_{W^\prime}^2+m_W^2-m_{\phi}^2)^2}{4m_{W^\prime}^2m_W^2}],\\
\Gamma(W_L^\prime\rightarrow H W)=\frac {g_{SM}^2}{24\pi m_{W^\prime}^2}p_f
[-6+\frac {(m_{W^\prime}^2+m_W^2-m_{\phi}^2)^2}{4m_{W^\prime}^2m_W^2}],\\
p_f =\frac{\sqrt{(m_{W^\prime}^2-(m_W+m_\phi)^2)(m_{W^\prime}^2-(m_W-m_\phi)^2)}}{2m_{W^\prime}},
\eeqn
 where the coupling is the same as the SM for $W_L^\prime$, $g_{SM}=k_+g^2/2$. It is Left-right symmetry for
 $W_L^\prime$ and $W_R^\prime$ in the above channel, while this symmetry is violated in the leptonic decay.
Because of the heavy mass of the right-handed neutrinos, the $W_R^\prime$ decay to leptons is not allowed .
 The leptonic decay width is only
\beqn
\Gamma(W_L^\prime\rightarrow l_i\nu_i)=\frac{m_{W^\prime}^2}{48\pi}g^2g_L^2.
\eeqn

\end{document}